\documentclass[10pt, a4paper,english]{article} 
\usepackage{csquotes}
\usepackage{float}
\usepackage{graphicx}

\usepackage{amsmath, amssymb, amsthm}
\usepackage{pictex, dcpic}
\usepackage{xcolor}

\def\al{\alpha}
\def\be{\beta}
\def\de{\delta}
\def\ga{\gamma}

\def\ep{\epsilon}
\def\io{\iota}
\def\te{\theta}
\def\la{\lambda}
\def\ze{\zeta}
\def\om{\omega}
\def\si{\sigma}
\def\vp{\varphi}

\def\ka{\kappa}

\def\Ga{\Gamma}

\def\La{\Lambda}

\def\Si{\Sigma}

 \def\calP{{\hbox{\cal P}}}

 \def\su{{\mathfrak{su}}}

 \def\C{\mathbb{C}}
 \def\D{\mathbb{D}}

 \def\G{\mathbb{G}}

 \def\K{\mathbb{K}}

 \def\P{\mathbb{P}}

\def\Aut{{\hbox{Aut}}}

\def\Spin{{\hbox{Spin}}}

\def\SU{{\hbox{SU}}}
\def\SL{{\hbox{SL}}}

\def\det{{\hbox{det}}}

\def\Diff{{\hbox{Diff}}}
\def\di{{\hbox{d}}}

\def\ip{\hbox to4pt{\leaders\hrule height0.3pt\hfill}\vbox to8pt{\leaders\vrule width0.3pt\vfill}\kern 2pt}

\def\del{\partial}
\def\na{\nabla}

\def\Lie{\hbox{\LieFont \$}}

\def\arr{\rightarrow}

\def\then{\Rightarrow}

\def\barJ{\bar J}

\def\calH{{{\cal H}}}

\def\calG{{{\cal G}}}
\def\calK{{{\cal K}}}
\def\calD{{{\cal D}}}

\def\frac[#1/#2]{\hbox{$#1\over#2$}}
\def\Frac[#1/#2]{{#1\over#2}}
\def\({\left(}
\def\){\right)}
\def\[{\left[}
\def\]{\right]}
\def\^#1{{}^{#1}_{\>\cdot}}
\def\_#1{{}_{#1}^{\>\cdot}}
\def\Label=#1{{\buildrel {\hbox{\fiveSerif \ShowLabel{#1}}}\over =}}
\def\<{\kern -1pt}

\def\circA{{}_\circ A}
\def\astz{{}^\ast\<\<z}
\def\circz{{}_\circ z}
\def\asth{{}^\ast\<\<h}
\def\circh{{}_\circ h}

\def\astk{{}^\ast k}

\def\astK{{}^\ast K}

\def\astL{{}^\ast L}

\def\aste{{}^\ast e}
\def\circe{{}_\circ e}
\def\astte{{}^\ast \te}
\def\circte{{}_\circ \te}
\def\astGa{{}^\ast \Ga}
\def\circGa{{}_\circ \Ga}
\def\Lie{\pounds}

\def\Uvec#1{\vbox{\mathsurround=0pt\ialign{##\crcr
     $\scriptscriptstyle\rightharpoonup$\crcr\noalign{\kern1pt\nointerlineskip}
     $\hfil\displaystyle{#1}\hfil$\crcr}}}
\def\Dvec#1{\vbox{\mathsurround=0pt\ialign{##\crcr
     $\scriptscriptstyle\rightharpoondown$\crcr\noalign{\kern-7pt\nointerlineskip}
     $\hfil\displaystyle{#1}\hfil$\crcr}}}

\def\uvecu{\vbox{\mathsurround=0pt\ialign{##\crcr
     $\scriptscriptstyle\rightharpoonup$\crcr\noalign{\kern1pt\nointerlineskip}
     $\hfil\displaystyle{u}\hfil$\crcr}}}
\def\dvecu{\vbox{\mathsurround=0pt\ialign{##\crcr
     $\scriptscriptstyle\rightharpoondown$\crcr\noalign{\kern-7pt\nointerlineskip}
     $\hfil\displaystyle{u}\hfil$\crcr}}}

\def\uvecbe{\vbox{\mathsurround=0pt\ialign{##\crcr
     \kern3pt$\scriptscriptstyle\rightharpoonup$\crcr\noalign{\kern1pt\nointerlineskip}
     $\hfil\displaystyle{\be}\hfil$\crcr}}}
\def\dvecbe{\vbox{\mathsurround=0pt\ialign{##\crcr
     \kern1pt$\scriptscriptstyle\rightharpoondown$\crcr\noalign{\kern-10pt\nointerlineskip}
     $\hfil\displaystyle{\be}\hfil$\crcr}}}

\def\uvecn{\vbox{\mathsurround=0pt\ialign{##\crcr
     $\scriptscriptstyle\rightharpoonup$\crcr\noalign{\kern1pt\nointerlineskip}
     $\hfil\displaystyle{n}\hfil$\crcr}}}
\def\dvecn{\vbox{\mathsurround=0pt\ialign{##\crcr
     $\scriptscriptstyle\rightharpoondown$\crcr\noalign{\kern-7pt\nointerlineskip}
     $\hfil\displaystyle{n}\hfil$\crcr}}}

\def\uvecm{\vbox{\mathsurround=0pt\ialign{##\crcr
     $\scriptscriptstyle\rightharpoonup$\crcr\noalign{\kern1pt\nointerlineskip}
     $\hfil\displaystyle{m}\hfil$\crcr}}}
\def\dvecm{\vbox{\mathsurround=0pt\ialign{##\crcr
     $\scriptscriptstyle\rightharpoondown$\crcr\noalign{\kern-7pt\nointerlineskip}
     $\hfil\displaystyle{m}\hfil$\crcr}}}

\def\uvecN{\vbox{\mathsurround=0pt\ialign{##\crcr
     \kern3pt$\scriptscriptstyle\rightharpoonup$\crcr\noalign{\kern1pt\nointerlineskip}
     $\hfil\displaystyle{N}\hfil$\crcr}}}
\def\dvecN{\vbox{\mathsurround=0pt\ialign{##\crcr
     \kern0pt$\scriptscriptstyle\rightharpoondown$\crcr\noalign{\kern-10pt\nointerlineskip}
     $\hfil\displaystyle{N}\hfil$\crcr}}}

\def\uvecu{\vbox{\mathsurround=0pt\ialign{##\crcr
     $\scriptscriptstyle\rightharpoonup$\crcr\noalign{\kern1pt\nointerlineskip}
     $\hfil\displaystyle{u}\hfil$\crcr}}}
\def\dvecu{\vbox{\mathsurround=0pt\ialign{##\crcr
     $\scriptscriptstyle\rightharpoondown$\crcr\noalign{\kern-7pt\nointerlineskip}
     $\hfil\displaystyle{u}\hfil$\crcr}}}

\def\uvecw{\vbox{\mathsurround=0pt\ialign{##\crcr
     $\scriptscriptstyle\rightharpoonup$\crcr\noalign{\kern1pt\nointerlineskip}
     $\hfil\displaystyle{w}\hfil$\crcr}}}
\def\dvecw{\vbox{\mathsurround=0pt\ialign{##\crcr
     $\scriptscriptstyle\rightharpoondown$\crcr\noalign{\kern-7pt\nointerlineskip}
     $\hfil\displaystyle{w}\hfil$\crcr}}}

\def\uvecv{\vbox{\mathsurround=0pt\ialign{##\crcr
     $\scriptscriptstyle\rightharpoonup$\crcr\noalign{\kern1pt\nointerlineskip}
     $\hfil\displaystyle{v}\hfil$\crcr}}}
\def\dvecv{\vbox{\mathsurround=0pt\ialign{##\crcr
     $\scriptscriptstyle\rightharpoondown$\crcr\noalign{\kern-7pt\nointerlineskip}
     $\hfil\displaystyle{v}\hfil$\crcr}}}

\def\barJ{\kern 3pt \bar {\kern -3pt J}}

\def\ShowLabel#1{\ref{#1}}

\def\ms{\medskip}
\def\ss{\smallskip}

\def\eq#1{\begin{equation}#1\end{equation}}
\def\eqLabel#1#2{\begin{equation}#1\label{#2}\end{equation}}

\def\Cases#1{\begin{cases}#1\end{cases}}
\def\Matrix#1{\begin{matrix}#1\end{matrix}}
\def\Align#1{\begin{aligned}#1\end{aligned}}

\def\eqs#1{\eq{\Align{#1}}}
\def\eqsLabel#1#2{\eq{\Align{#1}\label{#2}}}

\long\def\Note#1{\blockquote{\footnotesize #1}}

\date{}

\def\Figure[#1]#2{\begin{figure}[htbp] 
   \centering
   \includegraphics[#1]{#2} }

\def\EndFigure{\end{figure}}

\def\Diagram#1{\eq{
\begindc{\commdiag}[10]
#1
\enddc%
}%
}

\def\Item[#1]{\item[#1]}

\title{Lecture Notes in Loop Quantum Gravity.\\
LN3: Boundary equations for Ashtekar-Barbero-Immirzi model}

\author{\small  L.Fatibene$^{a,b}$,   A.Orizzonte$^{a}$\\
\\
\small$^a$ Department of Mathematics {\it``Giuseppe Peano''}, University of Torino (Italy)\\
\small$^b$ Ist. Naz. Fisica Nucleare (INFN) - Sezione Torino - Iniziativa spec. QGSKY (Italy)\\
}

\begin{document}


\maketitle


\begin{abstract}
We shall here perform the canonical analysis of field equations of ABI model in order to determine constraint equations.
We shall show that one can use algebraic constraints in the covariant framework to fix $k^i$ as a function of the frame and obtain a model where $(A^i_a, E_i^a)$ is a pair of independent fields 
which are also a pair of conjugated fields.

We shall not impose any relation on Immirzi parameter $\be$ and Holst parameter $\ga$, still constraint equations will depend on $\be$ only and they agree with standard result of LQG which are obtained by a suitable canonical transformation on a leaf of the ADM foliation used to define a Hamiltonian framework.

We eventually state the scheme for quantization that will be discussed in the following lecture notes.
\end{abstract}

\section{Introduction}

In LN1 we discussed the Ashtekar-Barbero-Immirzi (ABI) model (see \cite{LN1}). 
Here, we want to discuss its canonical analysis along the lines described in LN2; see  \cite{LN2}.

The ABI model is based on a further bundle reduction
\Diagram{
\obj(50,50)[beP]{$\Si$}
\obj(100,50)[P]{$P$}
\obj(150,50)[LM]{$L(M)$}
\obj(50,10)[M0]{$M$}
\obj(100,10)[M1]{$M$}
\obj(150,10)[M2]{$M$}
%
%
\mor{LM}{M2}{$\>_{GL(m)}$}
\mor{P}{M1}{$\>_{SL(2,C)}$}
\mor{beP}{M0}{$\>_{SU(2)}$}
\mor{beP}{P}{$\io$}[\atleft,\solidarrow]
\mor{P}{LM}{$e$}[\atleft,\solidarrow]
\mor{M1}{M2}{}[\atright,\solidline]
\mor(100,13)(150,13){}[\atright,\solidline]
\mor{M0}{M1}{}[\atright,\solidline]
\mor(50,13)(100,13){}[\atright,\solidline]
}
and on (a 1-parameter family of) fields $(e_I^\mu, A^i_\mu, k^i_\mu)$ defined on it.
The field $e$ is a spin frame on $P$, $A^i$ a $\SU(2)$-connection,
$k^i$ a 1-form valued in $\su(2)$, both on $[\Si\arr M]$.
The connection $A^i$ defines a covariant derivative which is denoted by $\na$ (for $\SU(2)$ objects), as well as a curvature 2-form $F^i$. 
The dynamics is described by the Lagrangian
\eqs{
L_H=& \frac[1/\ka ]      F^i \land  L_i   + \frac[1/\ka ]  \na k^{i} \land \( K_i-\be L_i\)   +\cr  &- \frac[1/2\ka] \ep_{ijk}  k^{i}\land   k^{j}\land \(  \(\be^2-1\)  L^k -2\be K^k \)   +   \frac[\La/3 \ka] \frac[\ga^2 /1-\ga^2]  K^k \land L_k 
}
where we set $K_k := \frac[1/2] \ep_{kij}  e^i\land e^j  - \frac[1/\ga] e^0\land e_k$ and $L_k :=   e^0\land e_k  + \frac[1/2\ga] \ep_{kij} e^i\land e^j $; see \cite{Rovelli2}.

This dynamics is $1st$ order with respect to the fields $A^i$ and $k^i$, it is $0th$ order with respect to the spin frame 
 (since $\na k^i$ does not depend on the Levi Civita connection).
It depends on two parameters, the {\it Holst parameter} $\ga$ which enters the action of the Holst model,  
and the {\it Immirzi} parameter $\be$ which parameterizes the reductive splittings used to define the connection $A$.
In the literature, one often takes $\ga=\be$; we leave them distinct here to show what is left of them in general; see \cite{HolstOriginal}, \cite{Barbero1},
\cite{Barbero2}, \cite{Immirzi}.

\Note{
Let us point out for later convenience that conjugate momenta to 1st order fields are given by
\eqs{
p_i^{\mu\nu}:=&\Frac[\del L_H/\del A^i_{\nu,\mu}] = \Frac[1/\ka ]    (e^0_\al e_{i\be}  + \frac[1/2\ga] \ep_{ijk} e^j_\al e^k_\be ) \ep^{\mu\nu\al\be}\cr
\pi_i^{\mu\nu} :=& \Frac[\del L_H/\del k^i_{\nu,\mu}] =  \Frac[1/\ka ] 
\(  \frac[\ga-\be/2\ga]  \ep_{ijk} e^j_\al e^k_\be -  \frac[1 + \be\ga/\ga]  e^0_\al  e_{i\be}  \)  \ep^{\mu\nu\al\be}
}
from which we see that assuming $\be=\ga$ would have quite an impact on the structure of momenta especially when projecting on the spatial foliation.

That also shows there are algebraic relations between tetrads and momenta which are a consequence of the fact that the tetrad enters the dynamics as a 0th order field.
}

The first 2 equations can be recast as 
\eq{
\Align{
\na (L^k+\ga K^k ) =& \ep^k{}_{lm}  \> k^l \land  \( K^{m}   - \ga  L^m\) 
- \be \ep^k{}_{lm}  \> k^l \land \( L^m   +\ga   K^m    \)
\cr
\na(K^k-\ga L^k) =& - \ep^k{}_{lm}  \> k^l \land \( L^m  +\ga  K^{m} \)
-  \be  \ep^k{}_{lm}  \> k^l \land \( K^m      -\ga    L^m  \)
}
}
The frame $e^I$ induces a spin connection $\tilde \Ga^{IJ}$ which can be split into $(\tilde A^i, \tilde k^i)$ and we can define the tensor objects
\eq{
z^i_\mu = A^i_\mu - \tilde A^i_\mu
\qquad\qquad
h^i_\mu = k^i_\mu - \tilde k^i_\mu
}
Since we know $\tilde \Ga^{IJ}$ is a torsionless connection by construction, we can write the first two equations as algebraic equations.

Field equations are
\eqLabel{
\Cases{
\na (L^k+\ga K^k ) = \ep^k{}_{lm}  \> k^l \land  \( K^{m}   - \ga  L^m\) 
- \be \ep^k{}_{lm}  \> k^l \land \( L^m   +\ga   K^m    \)
\cr
\na(K^k-\ga L^k) = - \ep^k{}_{lm}  \> k^l \land \( L^m  +\ga  K^{m} \)
-  \be  \ep^k{}_{lm}  \> k^l \land \( K^m      -\ga    L^m  \)     \cr
  F^k \land  e_k   -  \frac[1 +\be\ga/\ga]    \na k^{k} \land  e_k 
+  \frac[\ga-2\be-\ga\be^2/2\ga]  \ep^k{}_{ij} k^{i}\land   k^{j} \land  e_k   
+ \frac[\La/6]  \ep_{ijk } e^i\land e^j\land e^k
           =0\cr
  \( \ga F^h -  (1 +\be\ga)   \na k^{h} +  \frac[\ga-2\be-\ga\be^2/2]   \ep^h{}_{ij}  k^{i}\land   k^{j}  - \frac[\La/2]  \ep^h{}_{ij}  e^i\land e^j  \) \land  e^0  +\cr
\qquad+ \(  \ep^h{}_{kj}  F^k\land e^j   +  (\ga-\be)   \ep^h{}_{kj} \na k^{k}\land  e^j    + \( \be^2-1-2\be\ga \)  k^{h}\land    k_{l}\land   e^l  \) =0
}
}{CovariantFieldEq}

The main reasons to discuss the canonical analysis are to find constraint equations (which will later play a major role in LQG quantization) and which also provide us with algebraic relations among spatial fields which in fact traditionally have been imposed as {\it ad hoc} definitions, while here can be derived as a consequence of the spacetime action; see \cite{Orizzonte}, \cite{Orizzonte1}, \cite{NostroRov1}, \cite{NostroRov2}.
Although the computation is not too simple, we believe it is worth doing once and for all to show what is a definition and what can be proven as a consequence of the dynamics.
As we already mentioned, eventually we will be simply interested in the final constraint equations which are the basis of the following quantization procedure.

\section{Canonical analysis}

Let us fix an evolution bubble $(\bar D, \ze, i)$ and pull-back the configuration bundle $[C\arr M]$ on $D$ along the canonical embedding $\hat \imath: D\arr M$, which is simply a restriction on $D\subset M$.
Let $S$ be the spatial manifold (which is the set of all integral curves of $\ze$) on which rest bundle $[D\arr S]$ is defined;
see \cite{LN2}.

Let us set $i_t= \Phi_t \circ i:S\arr D $ for the {\it space leaf} at time $t$, where $\Phi_t$ is the flow of the evolution field $\ze$,
and $S_t$ for the images of $i_t$, which in fact produce a (regular) foliation of $D$.

We can  choose  coordinates $(t, k^a)$ adapted to the evolution field $\ze$ which hence in these coordinates reads as $\ze=\del_t$. 
Adapted coordinates also are adapted to the spatial foliation $S_t$, i.e.~the leaves read as $t=t_0$.

Then eventually we have a pre-quantum configuration bundle $[{\hat\imath}^\ast C\arr D\arr S]$, its sections being pre-quantum configurations.

\subsection{Adapted fields}

First thing is to adapt fields to the spatial foliation of the evolution bubble; see \cite{Gourgoulhon}.
To be precise we need a way to build covariant fields out of the spatial fields we are defining, both for pulling back the covariant dynamics on spacetime to space $S$ as well as
for eventually rebuilding covariant fields out of spatial solutions.
In general, for example for $A^i$ and $k^i$, we can define spatial fields by pulling back the covariant fields on the spatial leaf (namely setting $\astz^i := (i_t)^\ast z^i$ and $\asth^i := (i_t)^\ast k^i$)
or by first contracting them once along the transverse vector $e_0$ and then pulling back on the spatial leaf (namely setting  $\circz^i := (i_t)^\ast (e_0\ip z^i)$ and $\circh^i := (i_t)^\ast (e_0\ip h^i)$).

\Note{
Since we are using only fields which are locally described in terms of forms that is all.
Of course, the decomposition depends on the trivialization of the rest bundle $[D\arr S]$  {\it even if} the bundle is trivial.
For a $k$-form $\te^i$ valued in $\su(2)$,  we have ${4 \choose k}\times 3$ components.
When we split it, we have ${3 \choose k}\times 3$ for $\astte^i$ plus ${3 \choose k-1}\times 3$ for $\circte^i$.
Of course we have
\eqs{
&\( {3 \choose k} +  {3 \choose k-1} \) \times 3 
=\( \Frac[3! (3-k+1)/ k!(3-k+1)!] +  \Frac[3! k/ k!(3-k+1)!] \) \times 3 =\cr
&\quad=
 \Frac[3! \( 3-{k}+1 +  {k} \)/ k!(3-{k}+1)!]  \times 3 
= \Frac[4! / k!(4-{k})!]  \times 3 
={4 \choose k}\times 3
}
Hence the map $\te^i \mapsto (\astte^i, \circte^i)$ can be (and in fact it is when the evolution field is transverse to the leaves) one-to-one.
}

For the spin frame, that means we need a way to build a spacetime spin frame (namely a {\it tetrad}) out of a spatial spin frame (namely a {\it triad}).

Let us define {\it adapted tetrads}  $e_I=(e_0, e_i)$ which have $\ep^i := \aste^i$ tangent to the leaves $S_t$ on the evolution foliation, hence $n := e_0$ transverse to it.
We also have $\circe^i =0$, while for the cotedrad, we have $\aste^0 =0$ and $\circe^0=1$. 

\Note{
Not all tetrads are adapted, however one can always apply a gauge transformation $\Aut_V(P)$ to get an adapted tetrad. 
Since the original covariant dynamics is  $\Aut(P)$-covariant  that means that one can always find a {\it gauge-equivalent} adapted tetrad.
Of course, given an adapted tetrad, one is still free to act on it by $\Aut(\Si)$ (which acts fixing the vector $n=e_0$), to get tetrads which are still adapted.

It should be noticed that we are refraining from calling $e_0$ the normal vector, since we have no metric defined on $M$ although, given a tetrad, $e_0$ is normal with respect to the metric induced by that tetrad.
As a matter of fact,  any transverse vector $e_0= N^{-1}(\ze - \be^a\del_a)$ is normal for a suitable configuration and $e_0$, with its 4 degrees of freedom, is what one needs to rebuild the tetrad $e_I$ (16 dof) out of the triad 
$\ep_i$ (9 dof) plus the 3 dof representing the gauge freedom in $\Aut_V(P)$ to adapt frames, namely 6 dof of $\SL(2, \C)$ modulo the residual 3 dof in $\SU(2)$.

We parameterize $e_0$ by a (non-vanishing) {\it lapse function} $N$ and a {\it shift vector} $\uvecbe= \be^a\del_a$ on space $S$ as well as the triad with $\ep_i= \ep_i^a \del_a$.
Accordingly, we have a map $(N, \be^a, \ep_i^a)\mapsto (e_I^\mu)$ which is one-to-one on adapted tetrads, given by
\eq{
|e_I^\mu| = \(\Matrix{
e_0^0  	&		e_i^0  \cr
e_0^a  	&		e_i^a  \cr
}\)
= \(\Matrix{
N^{-1}	&		0  \cr
-N^{-1}\be^a 	&		\ep_i^a  \cr
}\)
\qquad
|e^I_\mu|  = \(\Matrix{
\bar e_0^0  	&	\bar e_a^0  \cr
\bar e_0^i  	&		\bar e_a^i  \cr
}\)
= \(\Matrix{
	N			&	0	\cr
	\be^a \ep^i_a 	&	\ep^i_a	\cr
}\)
}

A general tetrad is recovered if we add 3 pointwise boosts which are other 3 functions on space.

Let us remark that we have $e=\det(e)= N \det(\ep)= N\ep$.
Let us finally stress that by changing the lapse function or the shift vector one changes the foliation, i.e.~they are purely gauge dof, which in fact parameterize the conventions of an observer.
}


For later convenience let us introduce the {\it densitized triad} $E_i := \star \ep_i  = E_i^a dS_a$
\eq{
E_i = \star (\ep_i)  =  \ep_{ia} \star (dk^a) 
=\frac[1/2] \ep \ep_i^{a} \ep_{abc}d k^b\land dk^c
= \ep \ep_i^{a} dS_a 
=: E_i^{a} dS_a 
}
where $\star$ denotes the Hodge duality induced on $S$ by the induced metric (or the triad, which is the same thing) and $\ep$ is the determinant of the cotriad $\ep^i_a$.

\Note{
As a matter of fact, the field transformation $\ep^i_a \mapsto E_i^a$, although convenient, is just a field transformation.
There is also an equivalent and convenient way of expressing the densitized triad $E_i$ as a function of the triad, namely
\eq{
E_i = \frac[1/2] \ep \ep_i^{a} \ep_{abc}d k^b\land dk^c
=\frac[1/2]  \ep_{ijk} \ep^j_b \ep^k_c d k^b\land dk^c
=\frac[1/2]  \ep_{ijk} \ep^j \land  \ep^k 
}

As far as momenta $p_i^{\mu\nu}$ and $\pi_i^{\mu\nu}$  are concerned, we can split them with respect to the evolution foliation to get
\eq{
\Align{
\ka  p_i^{ab}=&N \ep_{ic}   \ep^{cab} + \frac[2/\ga]  E_i^{[a} \be^{b]}  \cr
\ka  \pi_i^{ab} =&    2\ka \frac[1+ \ga^2   /\ga^2]  p_i^{0[a}\be^{b]}   -  \frac[1 + \be\ga/\ga]    \ka \hat p^{ab}_i  
}
\qquad
\Align{
\ka p_i^{0a} =&     \frac[1/\ga]  E_i^a  \cr
\ka \pi_i^{0a} =&   \frac[\ga-\be/\ga]   E_i^a = \ka (\ga-\be)  p_i^{0a}
}
}
}

In view of the fact that $\aste^0= 0$, when we pull-back on $S_t$, we also have
\eq{
\astK_k = \frac[1/2] \ep_{kij}  \ep^i\land \ep^j 
=  \ga\> \astL_k =: \ga \> \astL_k = E_k
}

Let us remark that projecting (co)frames (as projecting metrics) is a complicated issue. Both the tetrad and the triad induce metrics $g$ on $M$ and $\ga$ on $S$, which induce connections $\tilde \Ga$ and $\tilde \ga$, which induce covariant derivatives $\tilde \na$  and $\tilde D$ (as well as curvatures tensors and extrinsic curvatures). 
If there was a thing we did not need is other sets of connections to deal with since we already have $\om^{IJ}$ and $A^i$.

We already know that on-shell (i.e.~along classical solutions) eventually the spin connection $\om^{IJ} = \tilde \Ga^{IJ}$ will agree with the one induced by the tetrad, however, we shall need to manipulate field equations before solving them.
We need to distinguish between identities that are general and identities which are satisfied only on-shell. This is the core of quantum formalism, to treat classical solutions as  approximation of what happens at a quantum level.
As a consequence, unfortunately, we need to keep all connections distinct, which is pretty annoying and confusing
even because some of them eventually will disappear on shell.
We need to do it, exactly because some will vanish because of evolution equations, some in view of constrained equations.
Later on in quantum formalism the point will be that we impose constraint equations on quantum states (because they are satisfied on physical states, the {\it physical state} being defined as what in fact identifies a solution), while evolution equations will be realized just by mean values,
i.e.~weakly up to quantum fluctuations.

\Note{
Let us then be patient and denote by $\tilde \Ga^{IK}_\mu = e^I_\al \(\{g\}^\al_{\be\mu} e^{J\be} + d_\mu e^{J\al}\)$ the spin connection induced by the tedrad, which satifies the property $\tilde \na_\mu e^I_\nu =0$.
This is pretty remarkable since it implies $de^I + \tilde \Ga^I{}_{J} \land e^J=0$, which in turn allows us to get rid of derivatives of the frame ($de^I =  e_J \land \tilde \Ga^{IJ} $) so to recast field equations 
involving derivatives of the frame as algebraic constraints.
Analogously, $\tilde \ga^{ij}_a$ denotes the connection induced by the triad.
We can split and project
\eqs{
\astGa^{0i} = &  \bar e^0_0 \(\{g\}^0_{\be a} e^{i\be} + d_a e^{i0}\) dk^a
=  N \eta^{ik}\{g\}^0_{b a} \ep_k^{b}  dk^a=\cr
=& -  \eta^{ik} \chi_{b a} \ep_k^{b}  dk^a
\cr
\astGa^{ij} =&  e^i_0 \{g\}^0_{bc} e^b_k \eta^{kj} dk^c +  e^i_a \(\{g\}^a_{bc} e^b_k +{ \del_c e^a_k}\)\eta^{kj} dk^c=\cr
=& 
 \ep^i_a \(\{\ga\}^a_{bc} \ep^b_k +{ \del_c \ep^a_k}\)\eta^{kj} dk^c= \tilde \ga^{ij}_a dk^a = \tilde \ga^{ij} \cr
\circGa^{0i} :=& \tilde \Ga^{0i}_\mu e^\mu_0  
= \eta^{ij}\{g\}^0_{b0} \ep_j^{b}  - \eta^{ij}\{g\}^0_{bc} \ep_j^{b}  \be^c=\cr
=&
 \eta^{ij} N^{-1}    \( D_b N - {{\be^a   \chi_{ba}} } +{{\be^a \chi_{ba}}}  \)\ep_j^{b} 
= N^{-1}    D_b N \ep^{ib}  \cr
\circGa^{ik} :=& \tilde \Ga^{ik}_\mu e^\mu_0 
= \(\ep^i_0 \{g\}^0_{b \mu} \ep^{kb} + \ep^i_a  \{g\}^a_{b \mu} \ep^{kb} +\ep^i_a d_\mu \ep^{ka}\) e_0^\mu =\cr 
=&N^{-1}\( \ep^i_a \( D_b\be^a -N \ga^{ad} \chi_{db}\)\ep^{kb} +\ep^i_a d_0 \ep^{ka} 
-\be^c \tilde \ga^{ik}_c \)=\cr
 =:& \ep^i_c   \de_n \ep^c_k   - \ep^i_c  \ga^{cd}\chi_{db} \ep^b_k\  \cr
}
where we used the identities
\eqs{
\{g\}^0_{bc}  =:  - N^{-1}\chi_{bc}
\qquad&
\{g\}^a_{bc} = \{\ga\}^a_{bc} + N^{-1} \be^a \chi_{bc} 
\cr
\{g\}^0_{b0}  = N^{-1}    D_b N - N^{-1}\be^a  \chi_{ba} 
\qquad&
\{g\}^a_{b0}  = -\be^a \{g\}^0_{b0} + D_b \be^a - N\ga^{ad} \chi_{db} 
}
which are tedious, though elementary, to be obtained. 
Here $\chi_{ab} := N^{-1} \(D_{(a} \be_{b)} -\frac[1/2] d_0 \ga_{ab} \)$ is the extrinsic curvature of $S$ in $M$, which is a symmetric bilinear form on $S$. (In LN2 (see \cite{LN2}) it was denoted by $K_{ab}$ but now we already used $K$ or $k$ for a lot of other objects.)
Notice that 
\eqs{
\astGa^{0i}  \land \ep_i  =&     -  \eta^{ik} \chi_{b a} \ep_k^{b}  \ep_{ic} dk^a\land dk^c
=       \chi_{ac}  \> dk^a\land dk^c =0\cr
\ep^i{}_{jk} \astGa^{0j}  \land \ep^k =&
-\ep^i{}_{jk} (   \eta^{jn} \chi_{b a}\ep^a_l \ep_n^{b} )\> \ep^l  \land \ep^k
= -\ep^{ink}    \chi_{nl}\> \ep^l  \land \ep_k =\cr
=& -\bar\ep \ep^{i}_d \ep^{dc}{}_b \chi_{ca}   \> dk^a \land dk^b
}

Let us also remark that $\astGa^{ij}$, $\circGa^{0i}$, as well as  $\astGa^{0[i}_a \ep^{j]a}$, do not depend on time derivatives of the triad fields.
}

Let us denote by $\na_\mu$ and $D_a$ the covariant derivatives induced by $A^i_\mu$ and $A^i_a$ on $M$ and $S$, as well as by $\hat \na$ the covariant derivatives induced by $\om^{IJ}$.

\Note{
Let us remark that on $M$ we can take covariant derivative $\na$ only of $SU(2)$-objects (i.e.~sections of bundles associated to $[\Si\arr M]$)
and that we shall use $\om^{IJ}$ and its covariant derivatives $\hat \na$ on $M$ only.
}

We are now ready to project field equations along the evolution bubble.
As for the fields, equations $E=0$ can be projected on the leaves ${}^\ast\<\<E=0$ or in the normal direction ${}^\ast\<\<(e_0\ip E)=0$).
We decide to write equations in terms of the projected fields $(N, \be^a, \ep^i_a, \asth^i, \circh^i, \astz^i, A^i_a)$, which are [37] fields (plus 3 dof to adapt the tetrads).

The first two equations can be projected to obtain the tangent parts [3]+[3] equations and the normal parts [9]+[9] to be further split in their symmetric and skew parts [3+6]+[3+6]. We obtain 
\eq{
\Align{
& \circh^i  = 0\cr
&  {}^\ast\<z^{(ab)}   =0\cr 
&  {}^\ast\<h_{[ab]}=0 \cr
}
\quad\Align{
& [3] \cr
& [6] \cr 
& [3] \cr
}
\qquad\qquad
\Align{
& 	{}_\circ z^k  =0	\cr
& 	  {}^\ast\<  h^{(ab)}  =0  \cr 
&	D_A E_i^A=0 \cr
}
\quad\Align{
& [3] \cr
& [6] \cr 
& [3]\qquad\hbox{\it (Gauss constraint)}\cr
}
}
The algebraic constraints fix the Immirzi form $k^i=\tilde k^i$ and normal part of the connection $\circA^i = {}_\circ \tilde A^i$
to be function of the triad (as well as $N$ and $\be$). 
We are left with the fields $(A^i_a, E_i^a)$ on the leaves and the Gauss constraint.

\ms
Then we are left with the other two field equations to be projected on $S$.
We have
\eqsLabel{
\Cases{
  F^k \land  \ep_k   -   \frac[1+\be\ga/\ga]   D k^{k} \land  \ep_k 
+  \frac[-2\be-\ga\be^2+ \ga/2\ga]  \ep^k{}_{ij} k^{i}\land   k^{j} \land  \ep_k   
+ \frac[\La/6]  \ep_{ijk } \ep^i\land \ep^j\land \ep^k
           =0\cr
   \ep^h{}_{kj}  F^k\land \ep^j   +  (\ga-\be)   \ep^h{}_{kj} D k^{k}\land  \ep^j    + \( \be^2-1-2\be\ga \)  k^{h}\land    k_{l}\land   \ep^l   =0
}
}{Field Equations 3e4}

From the first of these equation we have
\eq{
\( \ep^{ij}{}_{k}  F^k_{ab}   +2  ( \be^2 +1)      \astk^{i}_a   \astk^{j}_b     + \frac[\La/3] \ep^{ijk}\ep_{abc} E^c_k  \) E^{[a}_i E^{b]}_j    =0
}
which is called the {\it Hamiltonian constraint}, see Appendix for details.

From the second equation we obtain
\eq{
F^k_{ab}   E^b_k  =0
}
which is called the {\it momentum constraint}, see Appendix for details.

The normal parts of these equations [3]+[6], of which [3] are identically satisfied and the other [6] provide evolution equations, 
as in standard GR, 
for the induced metric $\ga_{AB}$.
We could predict this since field equations cannot determine the evolution of the triad which is in fact covariant with respect to pointwise spatial rotations which produce their hole argument.
Accordingly, only 6 degrees of freedom described by the triad can evolve deterministically, the other 3 are gauge freedom
and subject to a hole argument.

\subsection{Constraint equations}

Let us summarize the constraints, since LQG deals then with how to quantize these constraint equations, rather than with their classical origin.
We have $[18]$ fields $(A^i_a, E_k^a)$ and $[7]$ (Gauss, Momentum, and Hamiltonian) constraints
\eq{
\left\{\Align{
&DE_k=0\cr
&F^i_{ab} E_i^b=0\cr
&\( \ep^{ij}{}_{k}  F^k_{ab}   +2  ( \be^2 +1)      \hat k^{i}_a   \hat k^{j}_b     + \frac[\La/3] \ep^{ijk}\ep_{abc} E^c_k  \) E^{[a}_i E^{b]}_j =0
}
\right.
}
where we fix $\hat k^i := k^{i}_a dk^a = \astk^i = \astGa^{0i}$.
Notice that in order to be completely consistent, constraints cannot be saparated by evolution equations.

\subsection{Evolution equations}

Let us here count for evolution equations. We started with $[40]$ fields (namely $A^i_\mu$, $k^i_\mu$, $e^I_\mu$) and equations. 
Then we eliminated $[12]$ fields (namely $h^k$) using $[12]$ algebraic equations (namely $\asth^k=0$ and $\circh^k=0$),
as well as other $[3]$ fields (namely $\circz^k$) by using other $[3]$ algebraic constraints (namely $\circz^k=0$).
We have $[4]$ gauge fields ($N$, $\be^a$) to parameterize the foliations and 3 dof to adapt tetrads to the foliation.

We are then left with $[18]$ fields (namely $A^i_a$ and $E_k^a$) and $[40-12-3]=[25]$ equations. 
However, we have $[7]$ differential constraints ($[3]$ Gauss constraint, $[3]$ momentum constraint, $[1]$ Hamiltonian constraint).
Among these [18] equations, [6] are identically satisfied and the others [12] are evolution equations. 

Let us also remark that we have $[7]$ constraints, as we have $[4]$ gauge fields ($N$, $\be^a$) plus $[3]$ boosts to adapt frames to the foliation.
Accordingly, the constraint equations balance the fields with are left undetermined by evolution equations, which hence parameterize the pre-quantum state. 

By the way, the evolution equations we obtain are the same equations one obtains in standard GR, namely
\eq{
 \dot \chi_{ab} =  \>{}^3\<R_{ab} +    \chi \chi_{ab}  - 2    \chi_{ac} \chi^c{}_{b}  -\La\ga_{ab}
} 
As usual in standard GR, these equations are symmetric hyperbolic on the spatial leaves in harmonic coordinates and accordingly they uniquely determine the evolution of the induced metric up to spatial diffeomorphisms.

For initial conditions satisfying the constraints, one determines covariant solutions of the original Holst equations.
Uniqueness is lost because of covariance, existence is lost for constraint equation.
Accordingly, Holst model is at the same time over-determined and under-determined as far as one considers Cauchy problems, as usual.

Hence we expect the {\it state} of the system to be related to initial conditions satisfying constraint equations, up to $\Aut(\Si)$-gauge transformations.

\section{Hamilton-Jacobi equations}

In view of the canonical analysis, we have a theory in which the dynamical variables are $(A^i_a, E_i^a)$ which, by the way, are a pair of conjugate fields on $S$.
In view of quantization, it would be natural to consider a Hilbert space built from functionals of the connection and we want here to further support this assumption.

\Note{
We have to stress that {\it quantization procedure} is always ill defined, it always requires a quantum leap to jump from a classical to a quantum viewpoint.
As a matter of fact, one should define a quantum theory axiomatically and then investigate its classical limit.
However, we consider quantization as a list of motivations for the axiomatic definition of the quantum theory that we shall do. 
}

Moreover, we started from an action which is manifestly covariant with respect to $\Aut(\Si)$ and, by the standard hole argument, we know that, if we want the (classical) physical state to be deterministic,
then we are forced to identify the physical state with equivalence classes of sections modulo gauge transformations rather than with sections.
That is why in standard GR the gravitational field is not a single Riemannian manifold $(M, g)$ but better any diffeomorphic pair $[(M', g')]$ with $M'=\phi(M)$ and $g=\phi^\ast(g')$.
In other words, the gravitational field is not a Lorentzian metric, it is rather an equivalence class of Lorentzian metrics up to diffeomorphisms.

Similarly, here we have to investigate how gauge transformations in $\Aut(\Si)$ act on connections. 

\Note{
Let us remark that the bundle $[\Si\arr S]$ is a principal bundle with the gauge group $\SU(2)$. It hence has fibered coordinates $(k^a, U)$ with $U\in \SU(2)$; see \cite{Trautmann},  \cite{Gockeler}, \cite{book1}, \cite{book2}.
When we change coordinates, we have
\eq{
k'^a= k'^a(k)
\qquad\qquad
U' = \la(k)\cdot U
} 
for some local transition function $\la: U_{\al\be}\arr \SU(2)$.
On $\Si$, we have a {\it canonical right action} which is free, vertical, and transitive along the fibers, given by $R_{S}:\Si\arr \Si: (k, U)\mapsto (k, U\cdot S)$, which is well defined and independent of fibered coordinates.

Accordingly, we can define a right-invariant pointwise basis 
\eq{
\rho_i = \ep_{ij}{}^k U^j_n \Frac[\del/\del U_n^k]
\qquad\( TR_S \(\rho_i(p)\) = \rho_i (p\cdot S)\)
}
 of vertical vectors on $\Si$.

Let us consider a general flow of automorphisms $\Phi_s\in \Aut(\Si)$ on $\Si$ projecting on a flow of diffeomorphisms $\vp_s: S\arr S$ on the base manifold $S$, generated by a vector field 
\eq{
\Xi= \xi^a(x)\del_a+ \xi^i(x) \rho_i
= \xi^a(x)\( \del_a - A^i_a\rho_i\)+ \xi^i_{(V)}(x) \rho_i
}
where we set $\xi^i_{(V)}:= \xi^i + A^i_a \xi^a$.

Since we know how connections transform with respect to transformations in $\Aut(\Si)$, namely
\eq{
A'^i_a =  \barJ_a^b \(
\la^i_j  A^j_b
+\frac[1/2]\ep^i{}_j{}^l \la^j_m \di_b \bar \la^m_l\)
}
then the (right invariant) vector field $\Xi$ induces a vector field $\hat \Xi$ on the bundle of connections 
\eq{
\hat \Xi= \xi^a(x)\del_a- \(    d_a \xi^b A^i_b
-\ep^i{}_{jk} A^j_a \xi^k  + d_a  \xi^i  \) \Frac[\del / \del A^i_a]
}

On any bundle, we have a definition for the {\it Lie derivative} $\Lie_\Xi \si = T\si(\xi) - \Xi\circ \si$ of a section $\si$ along a vector field $\Xi$ (projecting over $\xi$)
which measures how much the section changes when dragged along the transformation $\Phi_s$.
By simply specializing to connections, we obtain directly
\eq{
\Lie_\Xi A^i_a :=\Lie_{\hat \Xi} A^i_a 
=\xi^b F^i_{ab} + D_a \xi^i_{(V)}
}
}

The Lie derivative of a functional $\Psi[A]$ is
\eqs{
\Lie_\Xi \Psi[A] =& \Frac[\de \Psi/\de A^i_a] \Lie_\Xi A^i_a
= \Frac[\de \Psi/\de A^i_a] (\xi^b F^i_{ab} + D_a \xi^i_{(V)})=\cr
=& \Frac[\de \Psi/\de A^i_a] \xi^b F^i_{ab} +  D_a\( \Frac[\de \Psi/\de A^i_a]  \xi^i_{(V)}\) + D_a \( \Frac[\de \Psi/\de A^i_a] \)\xi^i_{(V)}
}

Accordingly, a functional $\Psi[A]$ is invariant with respect to $\Aut(\Si)$ iff $\Lie_\Xi \Psi[A] =0$ for any (compactly supported) generator of automorphisms $\Xi$, i.e.~iff
\eq{
\Frac[\de \Psi/\de A^i_a]  F^i_{ab}=0
\qquad\qquad
D_a \( \Frac[\de \Psi/\de A^i_a] \) =0
}

These conditions are similar to the Gauss and momentum constraints. To obtain a precise correspondence, we need Hamilton-Jacobi (HJ) formulation of dynamics.
HJ equations can be derived within Hamiltonian formalism and are in fact (also in mechanics) equivalent to Hamilton, hence Lagrangian, dynamics.
However, the original insights was directly to show that the action functional along solutions can be written as a functional (or a function in mechanics) of the pre-quantum configuration only.

This is the foundation of HJ formalism, it solves the control problem for a general Hamiltonian system (given the initial and final positions, it finds the initial and final momenta to single out a classical solution starting from the initial position and arriving at the final position), which is the {\it classical propagator}. It is the basis of path integral quantization and it gives HJ equation which is the eikonal approximation of Schr\"odinger equation.
Thus, let us go this way which does not use directly all the symplectic machinery of Hamiltonian formalism (see \cite{Benenti}), which is very beautiful, simple (after all the definitions), canonical in mechanics, unfortunately not too canonical in field theory.

\subsection{Hamilton-Jacobi framework in a gauge natural field theory}

In mechanics the boundary of pre-quantum states is of dimension $m-1=0$, it is made of discrete points and functionals on pre-quantum states are in fact {\it functions}; see \cite{Newman}.
That cannot be the case in field theories in which the Hamilton principal function is expected still to be in fact a {\it functional} of the pre-quantum states, as we shall show below.

Let us consider a compact region $\bar D\subset M$,
a solution of field equations $\si:M\arr C: x\mapsto(x, y(x))$ as well as an infinitesimal symmetry $\Xi=\xi^\mu(x) \del_\mu + \xi^i(x, y) \del_i$ which drags solutions into solutions so that we can define the symmetry flow $\Phi_\ep: (x^\mu, y^i)\mapsto (x^\mu_\ep(x), Y^i_\ep(x, y))$ and
a 1-parameter family of solutions 
\eq{
\si_\ep= \Phi_\ep\circ \si\circ  \vp_{-\ep}: M\arr C:x\mapsto \(x, y^i_\ep(x)= Y^i_\ep(x_{-\ep}(x), y^i(x_{-\ep}(x)))\)
}

The symmetry flow $\Phi_\ep$ projects onto a flow $\vp_\ep:M\arr M: x^\mu \mapsto x^\mu_\ep(x)$ on spacetime.
We can drag the region $\bar D$ along the flow to define a 1-parameter family of embeddings $\io_\ep=\vp_\ep: \bar D\arr M: x^\mu \mapsto x^\mu_\ep = \vp^\mu_\ep ( x)$
and let us set $\bar D_\ep= i_\ep(\bar D)\subset M$. 

\Note{
By taking derivatives with respect to the parameter $\ep$, we have
\eq{
\xi^\mu (x) := \Frac[dx^\mu_\ep / d\ep]\Big|_{\ep=0} \kern-10pt (x)=: \de x^\mu(x)
\qquad
\xi^i (x, y) := \Frac[d Y^i_\ep / d\ep]\Big|_{\ep=0} \kern-10pt (x, y)
\qquad
\de y^i (x) := \xi^i (x, y(x)) 
}
which just set the relation between the flow and its infinitesimal generator.
We can also consider the infinitesimal generator of the family of solutions
\eq{
\Frac[d y^i_\ep / d\ep]\Big|_{\ep=0} \kern-10pt (x) = -\( -\xi^i(x, y(x)) +  d_\mu y^i(x) \xi^\mu(x)\) =: -\Lie_\Xi y^i 
} 
This generator can be prolonged to derivatives
\eq{
\Frac[d (y^i_\mu)_\ep / d\ep]\Big|_{\ep=0} \kern-10pt (x) = -d_\mu \Lie_\Xi y^i 
} 
}

When we write the change in the action functional, we can proceed by a change of integration variables which essentially means pulling back on $\bar D$
along $\io_\ep: \bar D\arr \bar D_\ep \subset M: x\mapsto x_\ep(x)$
\eqs{
\de S [\si] =& \Frac[d/ d \ep] \int_{\bar D_\ep=\io_\ep(\bar D)} \kern-25pt L(x, y_\ep(x), dy_\ep(x))d\si\Big|_{\ep=0}\kern-10pt
= \int_{\bar D}  \Frac[d/ d \ep]  \(  (\io_\ep)^\ast L\(x, y_\ep(x), dy_\ep(x)\)d\si \) \Big|_{\ep=0}\kern-10pt =\cr
=& \int_{\bar D}   \Frac[d/ d \ep] \( L(x_\ep, y_\ep(x_\ep), dy_\ep(x_\ep)) J d\si \) \Big|_{\ep=0}\kern-10pt =\cr
=&\int_{\bar D}  \( \Frac[\del \xi^\la/\del x^\la ]  L +\Frac[\del L/\del x^\la] \xi^\la
+ \Frac[\del L/\del y^i] (y^i_\la \xi^\la- \Lie_\Xi y^i )
+ \Frac[\del L/\del y^i_\al] \(    y^i_{\al\la} \xi^\la -d_\al  \Lie_\Xi y^i  \)
\) \> d\si   =\cr
=&\int_{\bar D}  \( d_\la (\xi^\la  L) 
- \Frac[\del L/\del y^i] \Lie_\Xi y^i 
- \Frac[\del L/\del y^i_\al] d_\al \Lie_\Xi y^i 
\) \> d\si  =  \cr
=&\int_{\bar D}  \( d_\al \(\xi^\al L - \Frac[\del L/\del y^i_\al] \Lie_\Xi y^i \)  
- {{\( \Frac[\del L/\del y^i]  - d_\al \Frac[\del L/\del y^i_\al] \)}} \Lie_\Xi y^i 
\) \> d\si  =\cr
=&\int_{\del D}  \( L\xi^\al - \Frac[\del L/\del y^i_\al] \Lie_\Xi y^i \)  \> d\si_\al =\cr
=& \int_{\del D}  \( -\(p_i^\al y^i_\be- L \de^\al_\be \) \xi^\be + p_i^\al \de y^i  \)  \> d\si_\al=\cr
=& \int_{\del D}  \( -\(p_i^\al y^i_\be- L \de^\al_\be \) \xi^\be + p_i^\al \de y^i  \) u_\al \> dS
}
where $J$ is the determinant of the Jacobian $\del_\al x^\mu_\ep$, $u_\al$ is the canonical covector associated to $\del D\subset M$ and $dS$ the local volume form on $\del D$.
We already get that, in general, the Hamilton functional is a {\it boundary functional} and that
\eq{
\Frac[\de S/\de y^i] = p_i^\al u_\al
}
defines the conjugate momenta which, in adapted coordinates, read as $ p_i^0$.
The Hamilton boundary functional may depend on all fields at the boundary or just some combinations, depending on the symmetries of the system.

\Note{
In mechanics, with a regular Lagrangian, there is not much to discuss. There is only one independent coordinate $t$, so Greek indices ``range'' on a single value $0$. Fields are identified with positions, first derivatives with velocities, and together with momenta are denoted by $(q^i, u^i, p_i)$.
Moreover, the boundary of the the region $D$ is discrete $\{t_1, t_0\}$ (or $t_1-t_0$ keeping orientation under account).

Hence we have
\eq{
\de S [\si] =  \(-\(p_i u^i- L  \) \de t + p_i \de q^i \) \Big|_{t=t_0}^{t=t_1}
=\(-\calH \de t + p_i \de q^i  \)  \Big|_{t=t_0}^{t=t_1}
}
where $\calH = p_i u^i- L $ is precisely the {\it total energy}, which is just the Hamiltonian written in terms of velocities rather than momenta.

Therefore, we learned that, in mechanics, the Hamilton function can be written as a function $S(t_0, q_0; t_1, q_1)$ of the initial and final positions, namely $q_0=q(t_0)$ and $q_1=q(t_1)$,
the initial and final momenta are given by
\eq{
p_0 = - \frac[\del S / \del q_0](t_0, q_0; t_1, q_1)
\qquad\qquad
p_1 = \frac[\del S / \del q_1](t_0, q_0; t_1, q_1)
}
as prescribed by the theory of generating functions of canonical flows, and 
\eq{
\frac[\del S / \del t] + H\(t, q,  \frac[\del S / \del q]\) =0
}
which is Hamilton-Jacobi equation, for both the initial and final position.

\ss
In field theory, the Hamilton functional is a  {\it functional} of the fields at the boundary.
For  example, for the Klein-Gordon case, we have $L= -\frac[\sqrt{g}/2] \(\vp_\mu\vp^\mu+ \mu^2\vp^2\) $
and consequently
\eq{
p^\mu :=\Frac[\del L/\del \vp_\mu ]= -\sqrt{g}\vp^\mu
\qquad\qquad
\hat T^\al{}_{\be} =  \sqrt{g} \vp^\al \vp_\be -\frac[\sqrt{g}/2] \( \vp^\mu\vp_\mu  + \mu^2  \vp^2 \)\de^\al_{\be}
}
Let us decompose the covariant momenta  $p^\mu =: \sqrt{g}\( \pi^a \del_a x^\mu + \pi u^\mu\)$
into their normal and tangent parts and set $\vp_a:=\vp_\mu \del_a x^\mu$,
so that we have
\eq{ 
\sqrt{g}\vp^\mu\vp_\mu = -p^\mu \vp_\mu
 = -\sqrt{g} \( \pi^a \del_a x^\mu \vp_\mu+ \pi u^\mu \vp_\mu\) 
 = -\sqrt{g}\( \pi^2 +\pi^a  \vp_a \)
}
and, consequently,
\eq{
u_\al \hat T^\al{}_{\be} =  \pi \sqrt{g}\vp_\be+\frac[\sqrt{g}/2] \(  \pi^2 +\pi^a  \vp_a - \mu^2 \vp^2 \) u_{\be}
}

From the Hamilton principal functional, we get
\eq{
\Frac[\de S/\de x^\be]=    u_\al   \hat T^\al{}_\be
\qquad
\Frac[\de S/\de \vp]=  u_\al p^\al = -\sqrt{g}\pi
}
While the second equation defines the conjugate momentum, the first can be split into the normal and tangential directions
\eq{
\Align{
&\Frac[\de S/\de x^\be] u^\be= \frac[\sqrt{g}/2] \(  \pi^2 -\pi^a  \vp_a + \mu^2 \vp^2 \)  \cr
&\Frac[\de S/\de x^\be] \del_a x^\be =   \pi \sqrt{g}\vp_a
=   -\Frac[\de S/\de \vp] \vp_a 
}
}
which are {\it Hamilton-Jacobi equations} for a Klein-Gordon field. The second Hamilton-Jacobi equation shows that the Hamilton funcional does not depend on the parameterization of the boundary, only on the boundary itself,
as well as, of course, on the boundary field $\vp$.

\ss
Let us consider, as a further example, Maxwell electromagnetism with the Lagrangian 
\eq{
L=-\frac[\sqrt{g}/4] F^{\mu\nu} F_{\mu\nu}
\qquad
p^{\al\be} := 
\Frac[\del L\> /\del d_\be A_\al]= \sqrt{g} F^{\al\be}
}
We can expand the variation of the Hamilton functional as
\eq{
\Frac[\de S/ \de A_\be]= -\sqrt{g} u_\al F^{\al\be}
\qquad
 \Frac[\de S/ \de x^\be] = \sqrt{g} u_\al \(F^{\al\mu} F_{\be\mu} -\frac[1/4] F^{\mu\nu}F_{\mu\nu} \de^\al_\be\)
}
In adapted coordinates, we see that $\Frac[\de S/ \de A_0]=0$, i.e.~the Hamilton functional does not depend on the field $A_0$, namely $A_0$ is a gauge field.
As for the second equation is concerned, we can split it into its normal and tangent component obtaining (in adapted coordinates)
\eqs{
& \Frac[\de S/ \de x^\ep] u^\ep=   \sqrt{g} \(F^{0b} F_{0b} +\frac[2/4] F^{0b}F_{0b} +\frac[1/4] F^{bc}F_{bc}  \)
=\frac[\sqrt{g}/2] \( E^{a}E_{a} + B^{a}B_{a}  \)
 \cr
& \Frac[\de S/ \de x^\ep] \del_a x^\ep =  \sqrt{g} F^{0b} F_{ab} 
= \sqrt{g} (E \times B)_a
} 
These are the {\it energy density} and the {\it Poynting vector} and they are again invariant with respect to changes of parameterizations on the boundary.
}

In the examples, we see that, depending on the symmetries of the system, the Hamilton functional can be independent of some of the fields at the boundary.

Our ABI model is basically a generally covariant gauge theory for the group $\SU(2)$, just with extra constraints.
Luckily enough, this kind of models have been studied extensively and many techniques have been developed to deal with them and their quantization.
Ironically, some of these techniques do not really work (the Hilbert space of quantum states turns out to be not separable).
However, they work in LQG due to an extra symmetry group the Hilbert space of quantum states turns out to be separable,
producing a well defined and meaningful quantum model.

In adapted coordinates $u = dt$, we have for ABI model
\eq{
\Frac[\de S/\de A^i_a] = p_i^{a \al } u_\al  =\Frac[1/\ka \ga] E_i^a
\quad\then  \ka \ga \> p_i^{a 0 }  = E_i^a
}
since along $\del D$ the only non-zero differential is $d\si_0$.

As a matter of fact, we shall consider the space of functionals $\Psi[A]$ of $\SU(2)$-connections on the principal bundle $[\Si\arr \del D]$
and we shall ask them to obey constraint equations, namely to satisfy
\eq{
\left\{\Align{
&  \na_a \(\Frac[\de \Psi/\de A^i_a]\) =0 \cr
&  F^i_{ab} \Frac[\de \Psi/\de A^i_a] =0\cr
&  \ep_{i}{}^{jk} \(F^i_{ab}+ 2(\be^2+1) k^j_{[a}k^k_{b]} +\frac[\La/3] \ep^{jkn} \ep_{abc}E_n^c\)  \Frac[\de \Psi/\de A^j_a]\Frac[\de \Psi/\de A^k_b  ]=0\cr
}
\right.
}
where $k^i_a = \tilde \ga^{0i}_a$ are written in terms of $E_i^a$.

We shall study bases of these Hilbert spaces and operators on them.
We shall also discuss operators which will be interpreted as observables of quantum geometry.

\subsection{Quantization scheme}

We are now ready to state our quantization scheme for LQG.
We shall start introducing the rigged Hilbert spaces $(\calK, \calK', \K)$ of {\it kinematical states}. These are made of functionals $\Psi[A]$ of $\SU(2)$-connections on the principal bundle $[\Si\arr \del D]$.

Then we shall impose constraints to single out invariant states.
This will define $SU(2)$-invariant states $(\calG, \calG', \G)$ first (which satisfy Gauss constraint)  and then $\Diff(S)$-invariant states in $(\calD, \calD', \D)$ which satisfy momentum constraint.
Bases of these rigged spaces will be {\it spin networks} and {\it spin knots}.

 As a matter of fact, spin knots will span a separable Hilbert space and we are able to discuss a number of operators to represent a number of observables of space geometry, which are observables of the gravitational fields.
 This will give us a definition of {\it quantum geometry}.
 
 Then one should implement the Hamiltonian constraint as well to define a further rigged Hilbert space $(\calP, \calP', \P)$ of {\it physical states}.
 However, for this last step, one quantizes directly the covariant action on spacetime to define {\it spin foams}.
 
Although eventually, spin foams are the fundamental description of spacetime geometries, namely of the gravitational field, defining in details spin networks and spin knots as well as the geometry operators on space is a good exercise and motivation to get ready for the final quantum framework.

\ 

\section{Conclusions and perspectives}

Starting from a covariant variational principle on spacetime for the ABI model, we provided a covariant setting for gravity which is dynamically equivalent to standard GR in dimension 4.
In this setting, we have an $\SU(2)$-connection $A^i$ well defined on the spacetime $M$ that, when later restricted to space $S$, becomes the usual real Barbero-Immirzi connection used in LQG.
Usually, the spatial connection is obtained by a canonical transformation on a spatial leaf $S$ within a Hamiltonian framework.
Here instead we have a spacetime counterpart that can be simply restricted to the spatial leaf.

Let us stress that the covariant $\SU(2)$-connection $A^i$ (which is defined on $\Si$) is not the restriction of the spin connection $\om^{IJ}$ (defined on $P$) to a subbundle $\Si\subset P$.
If it had to be, we could restrict only some spin connections $\om^{IJ}$, namely the ones which happen to be already tangent to the subbundle. 
As a consequence, we would have restrictions on the holonomy group (to be discussed).
Here, we are instead projecting the spin connection $\om^{IJ}$ on the $\SU(2)$-bundle, to a different Barbero-Immirzi projection $A^i$, the difference being parameterized by the $\SU(2)$-valued 1-form $k^i$.
As a result, we can project any spin connection, the map $\om^{IJ}\mapsto (A^i, k^i)$ is one-to-one, the holonomy of $\om^{IJ}$ is {\it encoded} into the holonomy of $A^i$, although the holomomies are not the same.

In the original framework we proposed, $(e^a, A^i, k^i)$ are independent fundamental fields and there is not much of a choice if one wants to define a generally covariant framework.
The generally covariant framework allowed us to analyse the canonical structure of the model on a purely Lagrangian viewpoint which, at least formally, is more homogeneous with what one would like to do later when discussing spin foams and path integral-like quantization. 

As a matter of fact, in this model, we found both differential and algebraic constraint equations.
As a first result, the covariant setting determines $k^i$ (as well as the normal part ${}_\circ A^i$) as a function of the densitized triad, so that we are not required to assume it by definition.
The field $k^i$ is determined by the algebraic constraint in the covariant ABI model.

We showed that what is traditionally done in LQG corresponds to considering the Holst model as a dynamical equivalent formulation of standard GR in dimension 4,
using the algebraic constraints to freeze the $\su(2)$ 1-form $k^i$ in it as a function of the frame and regarding the model as a functional of the connection.
The canonical analysis of this model provides constraint equations to be quantized for LQG.

Another result here is that we did not impose a relation between the Immirzi parameter $\be$ and the Holst parameter $\ga$.
These parameters have different origin. Immirzi parameter is kinematical, it parameterizes reductive pairs $(\SU(2), \SL(2, \C))$ which are used to define the projection of the spin connections $\om^{IJ}$ to the $\SU(2)$ connections $A^i$. The Holst parameter instead is dynamical, it parameterizes classically equivalent action principles.
We proved that even without fixing them, the constraint equations eventually depend only on the Immirzi parameter $\be$.
Let us remark that in a general dimension $m\not=4$ we have no Immirzi parameter, or better we are force to set $\be=0$.
On the other hand, in dimension different from $m\not=4$, we have no freedom in adding a Holst term to the standard action.
For example, in dimension $m=3$, one has the Lagrangian $L= R^{IJ}\land e^K\ep_{IJK}$, with no Holst parameter.

 Although we agree that our physical spacetime is in fact of dimension $4$, it may be interesting to explore what can be done in a general dimension, at least from a mathematical perspective. 
Let us also mention that, for a physical theory we need experiments that, at this stage, we do not have.
However, we can regard LQG at least as a proposal for how a general covariant field theory should look like when quantized, mathematically speaking.

Before proceeding to quantization, we still need to discretize connection kinematical functional. As a matter of fact using the connection components as variables is not a  good idea. Connection fields transform oddly with respect to gauge transformations and we eventually need to parameterize gauge classes of connection which, in view of the hole argument, are a better representation of physical states.

For these reasons, we need to introduce holonomies and show that they are in one-to-one correspondence with gauge classes of connections. Moreover, they transform reasonably with respect to gauge transformations.
Discretisation of connections will be instead done by introducing {\it film bundles} and their connections.
Both these issues should be understood in general, based on the theory of principal bundles, which is the natural arena for understanding these structures in full generality, including the old cases of differential geometries and holonomies of connections on a manifold.

We also need to review some basic facts about the gauge group $\SU(2)$, its irreducible representations, and the Haar measure on it which will play a major role in the definition of the scalar product of the Hilbert space of quantum states. This is also a key issue since we need to define the scalar product without using a metric on spacetime. For that, Haar measure is perfect since it is a canonical group structure; it exists on all compact groups (although we shall discuss it only on $\SU(2)$) and it is the reason why it was originally essential to define reduction from $\SL(2, \C)$ (which is not compact) to $\SU(2)$ (which is).

Let us remark that although Lie algebras are a key technique to investigate group representations, we here need to deal with {\it group} representations, not {\it Lie algebra} representations. In fact, we are going to apply the techniques to $SU(2)$-connections
which cannot be defined if not on a principal bundle, and principal bundles deal with {\it group} transformations, not to mention that gauge transformations are primarly {\it group} transformations.

The techniques presented in the following lecture notes are often borrowed from gauge lattice theories. Moreover, we know a great deal about specifically $\SU(2)$ since that is $\SU(2)\simeq \Spin(3)$, i.e.~the group one uses when dealing with angular momentum and spin.

\section*{Appendix A: Derivation of constraint equations}

Since we used the first two equations to express $k^i$ (and $\circA^i$) as a function of the triad we can express the (projections of the) curvature and the covariant derivatives of $k^i$ as
\eq{
\Align{
&{}^\ast\<F^i  =\frac[1/2] F^i_{ab}\> dx^a\land dk^b
\qquad\qquad\qquad\qquad\>\>
{}_\circ\<F^i  = F^i_a \>dk^a
\cr
&{}^\ast\<(\na k^i) =  {}^\ast\<(\na k)^i_{ba}\> dk^b\land dk^a 
\qquad\qquad\qquad
{}_\circ\<(\na k^i) ={}_\circ\<(\na k)^i_a\> dk^a
}
}
where we set
\eq{
\Align{
& \tilde F^i_{ab} = \frac[1/2]\ep^i{}_{jk} \ep^j_c \ep^{kd} \>{}^3\<R^{c}{}_{dab}
 +\be  \ep^{ic}  \(  D_b \chi _{c a}- D_a \chi _{c b}\)
  - \be^2 \ep^i{}_{jk} \ep^{jc} \ep^{kd}  \chi _{c a} \chi _{d b}		\cr
&F^i_a  :=  \bar N  \(  d_0 A^i_a -   \be^c F^i_{ca} +   \frac[1/2] \ep^i{}_{jk}  D_a \(d_0 \ep^j_c\ep^{kc} - \be^b D_b \ep^j_c\ep^{kc}   -D_c \be^b \ep^j_b \ep^{kc} \)  - \be D_a(\ep^{ib} D_b N)  \)\cr
& {}^\ast\<(\na k)^i_{ab} =    D_b \( \ep^{ic}\chi_{ca} \)\cr
& {}_\circ\<(\na k)^i_a = 
 -\ep^{id} \( \Lie_n \chi_{da}+ \chi_{ab} \chi^b{}_{a} \)
+\bar N\(  \ep^{i}_d  \be^c D_c \chi^d{}_{a}  -  D_a  \( \ep^{lc}D_c N  \) - \be\bar \ep \ep^i_c  \ep^{cdb}\chi_{da}  D_BbN  \)
}
}
Here $ \Lie_n \chi_{da}$ denotes the Lie derivative of the extrinsic curvature in the direction of $\vec n$.

We can check the identities
\eqs{
k^i_a k^j_b  E^a_{[i} E^b_{j]}  =&  \frac[1/2]\ep^2 \chi^i{}_m \chi^j{}_n   (\ep^m_a\ep^a_{i} \ep^n_b  \ep^b_{j} -\ep^m_a \ep^a_{j} \ep^n_b \ep^b_{i} )
=\cr
=&  \frac[1/2]\ep^2 \chi^i{}_m \chi^j{}_n   ( \de^m_i \de^n_j  - \de^m_j \de^n_i  )
=  \frac[1/2]\ep^2   (\chi^2  - \chi^i{}_j \chi^j{}_i  )
}
as well as
\eq{
\ep^{ijk}\ep_{abc} E^c_k  E^{[a}_i E^{b]}_j  
= \ep^3 \ep^{ijk}\ep_{abc} \ep^c_k  \ep^{a}_i \ep^{b}_j 
= \ep^3 \ep^{ijk}   \bar \ep \ep_{ijk} 
= 6 \ep^2 
}

We can write down the Hamiltonian constraint in terms of the extrinsic curvature as
\eq{
\Align{
 & \ep_k{}^{ij}  F^k_{ab}\ep^{a}_i \ep^{b}_j  
 +(\be^2+1) (\chi^i{}_a\chi^j{}_b - \chi^i{}_b \chi^j{}_a    )\ep^{a}_i \ep^{b}_j  
  - \frac[\La/3]\ep \ep^{ijk}\ep_{abc} \ep^c_k   \ep^{a}_i \ep^{b}_j  =\cr
&\qquad
  = \ep^{eab}  \(\frac[1/2]  \ep_{ecd} \>{}^3\<R^{cd}{}_{ab}
 +{2\be\bar \ep    D_b \chi _{e a}}
  - \be^2  \ep_{ecd}  \chi^c{} _{a} \chi^d{}_{b} \)    
 +(\be^2+1) \(\chi^2 - \chi_{ab} \chi^{ab}   \)
  - 2\La  =\cr
&\qquad
  = \>{}^3\<R
  - \be^2 ( \chi^2  -  \chi_{ab} \chi^{ab} )
 +(\be^2+1) \(\chi^2 - \chi_{ab} \chi^{ab}   \)
  - 2\La  =\cr
&\qquad
  = \>{}^3\<R + \chi^2 - \chi_{ab} \chi^{ab}    - 2\La  =0		\cr
}
}

\subsection{The Hamiltonian constraint}

The we can project the third equation
\eq{
\Align{
&  {}^\ast\<F_k\land  \ep^k -\frac[\be\ga+1/\ga]\> {}^\ast\<\<\(\na k^{k}\) \land  \ep_k  
 -  \frac[\ga\be^2+ 2\be-\ga/2\ga] \ep_{kij}  \>{}^\ast\<k^{i}\land  \>{}^\ast\<k^{j} \land  \ep^k = \frac[\La/6]  \ep_{ijk} \ep^i\land \ep^j \land \ep^k
\cr
&  \frac[1/2] \ep_k{}^{ij}  F^k_{ab} \ep\ep^a_i \ep^b_j
+\frac[\be\ga+1/\ga]     \( D_b \ep^{kd}  \ep_{kc}  \chi_{da}   + { D_b \chi_{ca} } \) \> \ep^{bac}   
 -  \frac[\ga\be^2+ 2\be-\ga/\ga]  \ep k^i_a k^j_b   \ep^a_{[i} \ep^b_{j]}   = \La \ep \cr
&   \frac[1/2]  \bar\ep  \ep_k{}^{ij}  F^k_{ab} E^a_i E^b_j
+\frac[\be^2\ga+\be/\ga] \bar \ep \ep^2  \( \chi^2  -   \chi_{da}  \chi^{ad}    \)    
 -  \frac[\ga\be^2+ 2\be-\ga/\ga]  \bar \ep k^i_a k^j_b  E^a_{[i} E^b_{j]}   
 = \frac[\La/6]\bar \ep  \ep^{ijk}\ep_{abc} E^c_k  E^{[a}_i E^{b]}_j 
\cr
&    \frac[1/2]  \ep_k{}^{ij}  F^k_{ab} E^a_i E^b_j
 + \frac[\ga(\be^2+1)/\ga]  k^i_a k^j_b  E^a_{[i} E^b_{j]}   - \frac[\La/6] \ep^{ijk}\ep_{abc} E^c_k  E^{[a}_i E^{b]}_j  =0
\cr
&    \(  \ep_k{}^{ij}  F^k_{ab}
 +2(\be^2+1) k^i_a k^j_b     - \frac[\La/3] \ep^{ijk}\ep_{abc} E^c_k  \) E^{[a}_i E^{b]}_j  =0
\cr
}
}

\subsection{Momentum constraint}

We can check the identities
\eq{
\Align{
 \ep^{i}{}_{kp} F^p\land \ep^k 
 =&   \frac[1/2]\ep^{i}{}_{kp} F^p_{ab}\ep^k_c \ep^{abc} d\si
 =    \frac[1/2]F^p_{ab} \ep\ep^{ie} \ep_{p}^f \ep_{cfe}   \ep^{abc} d\si=\cr
 =&   (F^p_{ab} E_{p}^a )   \ep^{ib}  d\si
 = -  \be\ep     D_d \(\chi^{de}-\ga^{de} \chi   \) \ga_{be}   \ep^{ib}  d\si
 }
}
hence
\eq{
F^p_{ab} E_{p}^a    =-  \be\ep     D_d \(\chi^{de}-\ga^{de} \chi   \) \ga_{be}
}

We can project the forth equation as
\eq{
\Align{
&  {}_\circ\<F_k\land  \ep^k 
=\frac[\be\ga+1/\ga] \>{}_\circ\<\<\na k^{k} \land  \ep_k  
+ { \frac[\ga\be^2+ 2\be-\ga/\ga] \ep_{kij}  \>{}_\circ\<k^{i}\land  {}^\ast\<k^{j} \land  \ep^k }
\cr
&\qquad\then\quad
F^k_a  \ep_{kb} \ep^{abc} 
= - \frac[\be\ga+1/\ga]  \( d_0 \chi_{ba}+ \chi_{be} \chi^e{}_{a} \)  \ep^{abc}  =0
\cr
&\qquad\then\quad
 \ep_{kB} F^k_a  \ep^{abc} 
 =  -\be \( d_0  \chi_{ba} +\chi_{bf}    \chi^f{}_a   \)  \ep^{abc} 
= 0 
\cr
}
}

\section*{Acknowledgements}

We also acknowledge the contribution of INFN (Iniziativa Specifica QGSKY and Iniziativa Specifica Euclid), the local research project {\it  Metodi Geometrici in Fisica Matematica e Applicazioni (2023)} of Dipartimento di Matematica of University of Torino (Italy). This paper is also supported by INdAM-GNFM.
We are also grateful to S.Speziale and C.Rovelli for comments.

L. Fatibene would like to acknowledge the hospitality and financial support of the Department of Applied Mathematics, University of Waterloo where part of this research was done.

\end{document}